\documentstyle[12pt]{article}

\def\be{\nopagebreak[3]\begin{equation}}
\newcommand{\ee}{\end{equation}}
\def\ba{\begin{array}}
\def\ea{\end{array}}

\def\inf{\infty}

\newcommand{\tr}{{\rm tr}\,}

\renewcommand{\d}{\partial}

\newcommand{\La}{\Lambda}
\newcommand{\la}{\lambda}
\newcommand{\ra}{\rightarrow}
\newcommand{\khi}{\chi}
\newcommand{\te}{\theta}
\newcommand{\om}{\omega}
\newcommand{\eps}{\varepsilon}

\begin{document}
\begin{titlepage}
\begin{flushright}
ENS-94/07,
March, 1994
\end{flushright}

\bigskip

\begin{center}
{\LARGE Principal Chiral Field at Large N}

\vskip 0.7truecm


 {V. A. Fateev
\footnote{On leave of absence from: Landau Institute for Theoretical Physics,
117940, ul. Kosygina 2, Moscow, Russia}\\
{\em Laboratoire de la Physique Math{\'e}matique
\footnote{Laboratoire associ{\'e} au CNRS URA-768}
,\\
Universite des Sciences et Techniques du Languedoc,\\
Place Eug{\`e}ne Bataillon, 34095  Montpellier Cedex, France } }
\vspace{1pc}

 {V. A. Kazakov
\footnote{On leave of absence from:
Academy of Sciences of Russia, Moscow, Russia}}\\
{\em Laboratoire de la Physique Th{\'e}orique de l'Ecole Normale
Sup{\'e}rieure
\footnote{Unit{\'e} propre du CNRS, associe{\'e} {\`a}
 l'Ecole Normale Sup{\'e}rieure et {\'a} l`Univ. Paris Sud }
,\\
24 rue Lhomond,
75231 Cedex 05,
France}
\vspace{1pc}

 {P. B. Wiegmann

{\em James Frank Institute, Enrico Fermi Institute and Department of Physics of
the  University of Chicago,
5640 S.Ellis Ave., Chicago ILL 60637, USA\\
and\\
Landau Institute for Theoretical Physics}}
\vspace{1pc}

{\large \bf Abstract}
\end{center}

We present the exact and explicit solution of the principal chiral field in
two dimensions for an infinitely large rank group manifold. The energy
of the ground state  is explicitly
found for the external Noether's fields of an arbitrary magnitude.
 The exact   Gell-Mann
-Low function  exhibits the asymptotic freedom behaviour at large
value
of the field in agreement  with perturbative calculations.
Coefficients of the perturbative expansion in the renormalized charge are
calculated. They grow factorially with the order showing the presence
of renormalons.
At small field we found an inverse logarithmic singularity
in the ground state energy at the mass gap which indicates that at
$N=\infty$ the spectrum of the theory contains extended objects rather then
 pointlike particles.

\vfill
\end{titlepage}

\section{Introduction}

Recent progress in understanding of lower dimensional string theories is
partially due to the advantage of discrete methods. It happened that the
matrix quantum mechanics, despite its simplicity and solvability
describes nonperturbative aspects of string theories at d=1 target
space, i.e. $c=1$ matter coupled to the two dimensional gravity.

The standard combinatorial methods  of matrix models, as well as
the continuous approaches have appeared so far to
be ineffective for higher dimensional target space.

  On the other hand, it has been
 known for a long time  that certain matrix
field
theories are completely integrable in 2d for an arbitrary size of the
 matrix
field. One of the most representative integrable matix field theories is
the {\it principal chiral field} (PCF) which describes a free field on a
principal , say $SU(N)$, manifold:
\be S = {N
\over 2 \la_0 } \int d^2x \  \tr [ \d_\mu g^{\dag} \d_\mu g  ]
\label{action} \ee
 where $g$ is an $N$x$N$ unitary matrix. Its large $N$ solution has been
 anticipated
for a long time to follow from its finite $N$ solution
\cite{PolWieg,W1,W2,ORW}.

Apart from being a model of Goldstone bosons
 whose strong interaction is entirely determined by the
 geometry of the
 manifold, this model has a long history of exploring its analogy to QCD
 \cite{Migdal} and  contour geometry of gauge fields \cite{Polyakov}.

The large N limit of the model is of particular interest. It is
conceivable that it describes a string theory in two physical
dimensions due to the analogy between planar Feynman graphs and the world
sheets of a string.
 Of course one should
not take this analogy literally: in asymptotically free theory neither
the coupling constant $\lambda_0$ nor a renormalized coupling  is  a
cosmological constant of a string  \cite{tHooft}: due to
renormalons the contribution of even planar graphs grows factorially with
the order.  However, some signs of the stringy behaviour will be seen
in the nonperturbative regime.

In this
paper we present the exact and   explicit large N solution for the
chiral field in two dimensions.

 The exact solution of the PCF
(both $S$-matrix and  Bethe Ansatz equations) was found in
Ref.\cite{PolWieg,W1}. It turns out  that the spectrum of the,
say,  $SU(N)$ model contains $N-1$ kinds of
massive particles. They form multiplets
belonging to the  diagonal of
the $SU(N)\otimes SU(N)$ associated with
 all fundamental
representations of the $SU(N)$ algebra, namely, the vector representation
and all  antisymmetric tensors according to the Dynkin diagram.
The spectrum of
 masses  is
\be
 m_l= m  {\sin({\pi \over N} l ) \over \sin({\pi \over N}  )  }
\label{mk}
\ee
where $l=1,...,N-1$ is the rank of a fundamental representation and
 $m=m_1$ is the mass of the vector particle.
In the two-loop approximation
it is
\be
m=  \Lambda { 1 \over \sqrt{\la_0} } exp({-4 \pi \over \la_0})
\label{mass}
\ee
where  $\Lambda$ is a cutoff.  All
particles are bound states of the vector particles.

At large $N$ we must distinguish two physically different situations:

 $N\rightarrow\inf$ but $m=m_1=fixed$. This means that $m_l=lm_1$, so that
the $l$-th particle is not a bound state any more. It is decomposed into $l$
vector particles. This suggests that the interaction vanishes in this limit
and we end up with a free massive field. This limit is not of a
particular interest.

Below we consider another and the only physically interesting limit:
 $N\ra \inf$ but the heaviest mass  $m_{N/2}=\mu$ of the largest
antisymmetric tensor remains fixed. In this case the masses fuse so that
the mass spectrum becomes continuous. The label running along the
Dynkin diagram becomes a continuous parameter. The energy of the massive
excitation with a momentum $p$ will be therefore
$\sqrt{p^2+m_l^2}\approx\sqrt{p^2+q^2}$ where we introduced
$q=\mu {\pi\over N}l$.
We observe that an extra dimension emerges from the matrix structure of the
field. This means that  at $N=\inf$ particles do not form a discrete
spectrum,
 so the theory ceases to be a theory of point like
particles.

We shall find the energy of the ground state as a
function of "the Noether's"  field by adding a term
$tr (H_L Q_L+H_R Q_R)/2$
to the hamiltonian of the theory corresponding to the lagrangian (1).
Here
$Q_L= \int d^2x g \d_0 g^{-1} $ and
$Q_R= \int d^2x g^{-1} \d_0 g $ are
the Noether's left and right charges and $H_{R(L)}= diag(h_1, h_2-h_1, ...,
h_{N-1}-h_{N-2}, -h_{N-1})$ is an element of the Cartan subalgebra.
It amounts to introducing in eq.(\ref{action}) the covariant derivative
\be
D_\mu g = \d_\mu g -i  \delta_{\mu0} (H_L g+g H_R)/2
\label{cov}
\ee
instead of the usual derivative $\d_\mu g$. In what follows, we shall
consider only the case $H_L=H_R=H$.

 Parameters $h_i$ play the role of chemical potentials for elementary particles
of the model, so that the energy ${\cal E}(h)$ is the energy of the ground
state with a symmetry of the  Young tableau $[1^{{\cal N}
 -{\cal N}_1},2^{{\cal N}_1-{\cal N}_2},...N^{{\cal N}_{N-1}}]$
where ${\cal N}_l=-d/dh_l{\cal E}(h)$. The field $H$ introduces an
energy scale
into the theory and gives a valuable physical information.

The large $N$
solution is  explicit. To ease the references we state it now.

We use a special direction of the field which
excites all types of the particles on equal footing:
 \be
h_l= h { \sin({\pi \over N} l ) \over \sin({\pi \over N}  )  }
\label{hk}
\ee
 We show that the energy of the ground state is expressed in terms of
modified Bessel
functions:
\be
 f(h) \equiv {1\over N^2}
\Big({\cal E}(h)-{\cal E}(0)\Big) = - {h^2 \over 8\pi} B^2 I_1(B) K_1(B)
\label{fh}
\ee
where the parameter $B$ is defined through
\be
{m \over h} = B K_1(B)
\label{mh}
\ee

The distribution of rapidities of physical particles will obey the simple
semi-circle law with the support $B$. The parameter $B$ defines the value of
rapidity corresponding to
the Fermi momentum of the fused particles.   We shall
see, that  $B$ gives the most natural definition of the renormalized (
running) coupling constant :
\be
 {\bar \la}(h)= {4\pi \over
B}
\label{la}
\ee
 The reader may find some
similarity between $B$ and  the Fermi level of
 eigenvalues of the matrix field in the c=1 string theory
\cite{Kaz}.

The paper is organised as follows:

 In  section 2 we review
the derivation of the spectral Bethe ansatz integral equations for
rapidities.

In section 3 we solve the spectral equations in the large N limit.

In  section 4 we  obtain a singular
behaviour on the treshold $h \sim m$ of the spectrum.

In  section 5 we show that our solution agrees with perturbative two loops
calculations
at large $h/m$  and find exact value of the mass ( namely the ratio
$m/\La_{\bar {MS}}$).

In section 6 we calculate  all terms of perturbation theory in
running coupling constant
 $\bar \la$ and show their factorial growth.

\section{$S$-matrix and  Bethe-ansatz equations for any $N$}

Perhaps the most economical way to obtain the  Bethe-ansatz
equations for the chiral field
 is the factorized bootstrap method \cite{W2}, rather than direct
diagonalization of the hamiltonian of the model \cite{PolWieg,W1}. The point is
that the $S$-matrix can be easily found on the basis of some heuristic
hypothesis. Let us give a sketch of this approach.

1){\it $S$-matrix}. The chiral field is renormalizable and asymptotically
 free
\cite{Polyakov2},\cite{stone}.  It is invariant under
the left-hand  and right-hand group transformations $G\otimes G$ and
the action (\ref{action})
is defined only by the Lie algebra of $G$, i.e. does not depend on the
representation of the group $G$. Therefore it is natural to assume that
the elementary particles are massive and belong to some fundamental
representations of the diagonal of the $G\otimes G$, whereas the
antiparticles form conjugated representations. The model is integrable
\cite{Witten}, therefore the scattering is factorized. Under these
assumptions the minimal $S$-matrix (factorized scattering matrix with a
minimal set of singularities) can be determined unambiguously. The
complete proof of the minimal $S$-matrix being the scattering matrix of
the chiral field requires a more sophisticated technique (see
e.g.\cite{PolWieg,W1}).

It turns out that once we assume that there is a particle in some, say,
$l$-th, fundamental representation,  the factorized bootstrap tells us that
there are particles in all $N-1$ fundamental representations. In fact,
they are  bound states of an arbitrary chosen representation.
Therefore it is convenient to start from the vector particle. The
factorized $SU(N)\otimes SU(N)$ scattering matrix for vector particles
is the tensor product of the $SU(N)$ factorized vector $S$-matrices
 ${\cal S}=X(\theta) S(\theta) \otimes S(\theta)$.
Here $\theta$ is a rapidity
of  a massive relativistic particle
($p^0=m\cosh\theta$,\,$p^1=m\sinh\theta$) and $X(\theta)$  is the
CDD-ambiguity factor which cannot be determined by the factorization,
unitary and crossing symmetry conditions. The $SU(N)$ unitary, crossing
invariant,  factorized $S$-matrix of vector particles is well known
\cite{karowski}. It is
\be
 S(\theta)=u(\theta)(P^{+}+{\theta +i2\pi/N\over \theta-i2\pi/N}P^{-})
\label{S}
\ee
 where $P^{\pm}$ is  the projection operator onto symmetric
(antisymmetric) states.

 The
amplitude in the symmetric channel $u(\te)$ and the amplitude in the
cross channel (particle- antiparticle scattering) $t(\te)=
{1/2-\te/(2i\pi) \over 1/2-1/N-\te/(2i\pi)} u(i\pi-\te)$ obey the unitarity
 conditions $t(\te)t(-\te)=u(\te)u(-\te)=1$. The minimal solution of
these equations is
  \be u(\theta)={{\Gamma
(1-{{\theta}\over{2\pi i}})\Gamma ({{1}\over{N}}+{{\theta}\over{2\pi
i}})}\over{\Gamma (1+{{\theta}\over{2\pi i}})\Gamma
({{1}\over{N}}-{{\theta}\over{2\pi i}})}} \label{u} \ee
 Finaly the CDD-factor is chosen
to cancel all double zeros and double poles on the physical sheet
$0<Im\theta<\pi$:
\be
X(\theta)={\sinh({\theta\over 2}+{i\pi\over N})
\over \sinh({\theta\over 2}-{i\pi\over N})}
\label{S2}
\ee
This is the $S$-matrix of the vector particles. It has a pole on the
physical sheet at $\theta_b=2\pi i/N$ in the antisymmetric channel. It
corresponds to the first bound state ( the second rank antisymmetric
tensor) with a mass $m_2=m\sin(2\pi /N)/\sin(\pi/N)$. The $S$-matrix of these
particles can be also found by tensoring the vector $S$-matrix
(the fusion procedure). It also has a pole in the antisymmetric channel, and so
on. In this way   the whole mass spectrum (\ref{mk}) can be generated.

2){\it Bethe-Ansatz Equations}. The thermodynamical properties of the model can
be obtained by imposing the  periodic boundary conditions. For an integrable
problem they imply the balance of two particle scattering phases and a phase
of a free motion between collisions. For the $i$-th particle with
the momentum $m\sinh\theta_i$, the periodic boundary conditions lead to the
problem of the diagonalization of a product of two-particle S-matrices. Say,
for a state with $\cal N$ vector particles in the box $L$ we have
\be
\exp(imL\sinh \theta_\alpha)=\prod^{\cal N}_{\beta=1,\beta\ne\alpha}\cal
{S}_{\alpha \beta}(\theta_{\alpha}-\theta_{\beta})
\label{ba}
\ee
The eigenvalues
of the operator in the r.h.s of the eq.(\ref{ba}) can be found by the
projection onto the symmetric subspace
\be
\exp(imL\sinh \theta_{\alpha})=\prod^{\cal
N}_{\beta=1,\alpha\ne\beta} exp(i\phi(\theta_{\alpha}-\theta_{\beta}))
\label{ba1}
\ee
 where $exp(i\phi(\theta)) =u^2(\theta)X(\theta)$ .
To obtain the Bethe-Ansatz equation for the state which contains all kinds of
particles, say, for the state with the  Young tableau $[1^{{\cal N}
 -{\cal N}_1},2^{{\cal N}_1-{\cal N}_2},...N^{{\cal N}_{N-1}}]$ one has to
consider complex rapidities of the bound states -"strings"
$\theta\rightarrow\theta^l+2r\pi i/N$, where $\theta^l$ is a rapidity of the
$l$-th particle and $r$ is an integer running between $-l/2$ and $l/2$.
Substituting this into  eq.(\ref{ba1}) and multiplying equations over $r$
we shall obtain the  equations for the rapidities of the state which contains
${\cal N}_l$  particles of the kind $l$:
\be
\exp(iLm_l \sinh \theta^{(l)}_{\alpha})=\prod _{n=1}^{N-1}\prod^{{\cal
N}_n}_{\alpha=1,\beta\ne\alpha}exp(i\phi_{ln}(\theta_{\alpha}^{(l)}-\theta_{\beta}^{(n)}))
\label{ba2}
\ee
where
\be
\phi_{ln}(\theta)=
\sum_{-l/2<r<l/2,-n/2<r^{\prime}<n/2}\phi(\theta+2r\pi/N+2r^{\prime}\pi/N)
\label{phase}
\ee
After tedious calculations \cite{W1} we obtain
\be
{d\phi_{ln}(\theta) \over  d\theta}=
-2\int_0^{\inf} d\omega [R_{ln}(\omega)-\delta_{ln}] \cos\omega\theta
\label{phi}
\ee
where
\be
R_{ln}(\om)=2{{\sinh\Big(\pi\om(1-{1\over N}max(l,n))\Big)
\sinh\Big(\pi \om {1\over
N} min(l,n)\Big)}\over \sinh\pi\om }
\label{rln}
\ee
Taking  logarithm of the both sides of the Eqs.(\ref{ba2}) we obtain the
Bethe -Ansatz equations
\be
m_l\sinh \theta^{(l)}_{\alpha}={{2\pi}\over L}J^{(l)}_{\alpha}+
{1\over L}\sum _{n=1}^{N-1}\sum^{{\cal
N}_n}_{\alpha=1,\beta\ne\alpha}\phi_{ln}(\theta_{\alpha}^{(l)}
-\theta_{\beta}^{(n)})
\label{ba3}
\ee
where integers $J$ are the quantum numbers of the states.
 For the generalization of these equations to an arbitrary
Bethe states see \cite{W1,ORW}.
The energy of this state is obviously
\be
E={1 \over L} \sum_{l=1}^{N-1}
m_l \sum_{\alpha=1}^{{\cal N}_l} \cosh\te_\alpha^{(l)}
\label{energy}
\ee

3) {\it Spectral Equations}. The next step is to find rapidities to
minimize the energy (\ref{energy}) in the thermodynamic limit
${\cal N}_l/L=n_l$, while $L\rightarrow \inf$. We assume that the minimum of
the energy corresponds to a dense smooth set of $\te$'s, so one can
describe them by the distribution function of rapidities of particles
$\rho_l(\te)$ and the distribution of holes $\tilde\rho_l(\te)$. They are
related by the equation
\be
{1\over 2\pi} m_l\cosh
\theta=\tilde\rho_l(\te)
+\sum_n\int d\te' R_{ln}(\te-\te')\rho_n(\te')
\label{ro}
\ee
 where
\be
 R_{ln}(\theta)=
{1\over \pi} \int_0^{\inf}d\omega R_{ln}(\omega)\cos\omega\theta
\label{R}
\ee

Let us now turn to the energy of the  state at the given fields $h_l$
\be
{\cal E}=min_{n_l}\big(E-\sum h_ln_l\big)=\sum_l\int d\te
[m_lcosh\te-h_l]\rho_l
\label{energy1}
\ee
Consider a small variation of $\rho_l(\te)$ and $\tilde\rho_l(\te)$  and
introduce the (pseudo)energy
$\eps^{+}_l(\te)>0$  of a hole and the (pseudo)energy $\eps^{-}_l(\te)<0$
of a particle \cite{JNW}, such as
$\delta{\cal E}
=\sum_l\int d\te [\eps^{+}_l(\te)\delta{\tilde
\rho}_l(\te)-\eps^{-}_l(\te)\delta\rho_l(\te)]$.
 According to this definition at the
ground state
$\eps^{+}_l(\te)=0$ if $\tilde\rho_l(\te)\ne 0$ and
$\eps^{-}_l(\te)=0$ if $\rho_l(\te)\ne 0$, i.e.  $\eps^{+}_l (\eps^{-}_l)$ and
$\tilde\rho_l(\rho_l)$ have nonoverlapping supports.
Comparing with (\ref{energy1})
and using  (\ref{ro}) we find the spectral equations of the model \cite{W1}
\be
  h_l -m_l\cosh\theta=\eps^{-}_l(\te)+\sum_n\int
  R_{ln}(\te-\te^{\prime})\eps^{+}_n(\te^{\prime}),
\label{inteq}
\ee
\be
  {\cal E} =-{1\over 2\pi}\sum_l \int \eps^{+}_l(\te)m_l \cosh\te,
\label{egs}
\ee
  where $\eps^{+}_l(\te)>0,\,\eps_l^{-}(\te)<0, \
  \eps^{+}_l(\te)\eps_l^{-}(\te)=0$.

4){\it Diagonalization of the Spectral Equations}.
To prepare the spectral equation (\ref{inteq})  for
the large N limit, we diagonalize the  kernel matrix $R_{ln}$.
This is easy to do
since it reflects the structure of the Dynkin diagram \cite {ORW}. Indeed, its
inverse is
\be
R^{-1}_{ln}(\om) =
 \sinh^{-1}{\pi |\om| \over N}
\Big(\delta_{ln} \cosh{\pi \om \over N} -
1/2(\delta_{l,n+1} + \delta_{l+1,n})\Big)
\label{kernel}
\ee
Let us introduce orthonormal eigenfunctions of the $(N-1)$x$(N-1)$
Cartan matrix $C_{jk} = 2 \delta_{jk} -\delta_{j,k+1}- \delta_{j+1,k}$:
\be
\ba{rcl}
&& \khi^{(p)}_j =
\sqrt{2/N} \sin{\pi p j \over N}, \ \ \ \ p=1,2,...,N-1\nonumber   \\
&&   \sum_{k=1}^{N-1} \khi^{(p)}_k \khi^{(p')}_k = \delta_{pp`}
\ea
\label{Cart}
\ee
They are also the eigenvectors of $R_{ij}$
\be
R_{jk}(\om) =
\sum_{p=1}^{N-1} \khi^{(p)}_j \khi^{(p)}_k R^{(p)}(\om)
\label{kerdi}
\ee
where
\be
R^{(p)}(\om)=
 {\sinh{\pi |\om| \over N} \over  \cosh{\pi \om \over N} -
\cos{\pi p \over N} }
 = {2 N \over \pi} \sum_{r=-\infty}^{\infty}
 {|\om| \over \om^2 + (p+r N)^2}
\label{kerp}
\ee
Then the spectral equations can be diagonalized
with respect to the particle indices:
\be
\eps^{(p)}_{-}(\te)+\int
R^{(p)}(\te-\te^{\prime}) \eps^{(p)}_{+}(\te^{\prime}) {d\te^{\prime}} =
h^{(p)} - \delta_{p,1} M \cosh \te,
\label{equp}
\ee
\be
   {\cal E}=-{m\sqrt{N/2}\over 2\pi \sin(\pi/N) }
\int d\te \cosh\te\eps_{+}^{(1)}(\te)
\ra -{ m (N/2)^{3/2}\over \pi^2 }\int d\te \cosh\te\eps_{+}^{(1)}(\te)
\label{grst}
\ee
 where
\be
 M=\sum_{k=1}^{N-1} \khi_k^{(1)}
m_k={m \sqrt{N} \over \sqrt{2} \sin(\pi/N)}
\ra  {m N^{3/2} \over \sqrt{2} \pi}
\label{MM}
\ee
and
\be
\eps_{\pm}^{(p)}\equiv \sum_{k=1}^{N-1} \khi^{(p)}_k \eps_k^{(\pm)}; \ \ \ \
 h^{(p)}\equiv \sum_{k=1}^{N-1} \khi^{(p)}_k h_k
\label{ephp}
\ee
It is important to note that this transformation is valid and
eq.(\ref{equp}) holds, providing that linear
combination of particle's (hole's) (pseudo)energies
remain positive (negative): $\eps_{(+)}^{(p)}>0\,(\eps_{(-)}^{(p)}<0 )$.

\section{Large $N$ solution}

One has to be careful using the last definition of the Fourrier
transform of $\eps^{(p)}_{\pm}(\om) = \int_{-\infty}^\infty d\te
\eps^{(p)}_{\pm}(\te) \cos(\om \te)$: different $\eps_k^{+}(\te)$
entering the sum in (\ref{ephp}) have
different supports $(-B_k,B_k)$.
It has to be taken into account when solving
the integral equations.

However, if we take $h_k$ in the form (\ref{hk}), so that
\be
h^{(p)} = h \delta_{p,1} {\sqrt{N} \over \sqrt{2} \sin(\pi/N)}
\ra_{N
\ra \infty} {N^{3/2} \over \sqrt{2} \pi} h \delta_{p,1}
\label{h1}
\ee
it turns out that the ansatz
\be
\eps^{(p)}_{\pm}(\om) = \eps_{\pm} (\om) \delta_{p,1}{\pi \over \sqrt{8 N}
\sin(\pi/N)}
\ra_{N\ra \infty} {N^{1/2} \over \sqrt{8} } \eps_{\pm}(\om) \delta_{p,1}
\label{e1}
\ee
satisfies all the equations (\ref{equp}) for $p=2,3,...,N-1$
and the equation with $p=1$ (the Perron-Frobenius mode) gives an
integral equation for the definition of $\eps_{\pm}(\om)$. Since
$\eps_{+}(\te)$ is strictly positive it can be
viewed as a newly defined density, with all appropriate analytical
properties.

 It implies that we look for a solution of (\ref{inteq})
with equal supports $B_1=B_2=...=B_{N-1}=B$. In what follows we
consider  eq.(\ref{equp}) only inside the interval $(-B,B)$
where $\eps_-(\te)=0, \ \ \eps_+(\te) \equiv \eps(\te) >0$.
It is the only function  which contributes to the
ground state energy (\ref{grst}). It obeys  eq.(29) with $p=1$.

Further simplifications occur in the large $N$ limit.
The kernels $R^{(p)}$ look as:
\be
R^{(p)}(\om)=
{2N \over \pi} {|\om| \over   \om^2 + p^2 } , \ \ \ \ N\ra\infty
\label{kerN}
\ee
Finally, for the choice (\ref{h1}) of the  field we obtain the
integral equation:
\be
\int_{-\infty}^\infty {d\om\over 2\pi} \cos(\te \om) {|\om| \over \om^2
+1}\eps(\om) = h- m \cosh \te, \ \ \ \ |\te| \le B
\label{infNq}
\ee
where
$$ \eps(\om)=\int_{-B}^B {d\te} \eps(\te) \cos(\te \om)$$
Note that, in the large $N$ limit $R^{(p)}(\om)$ vanishes at large $\om$,
 whereas
at finite $N$ it approaches  $1$ (see eq.(28)). This implies that $\eps(\te)$
is no longer differentiable at $\te=\pm B$, but has a cusp. As a result the
physics on the threshold $h\sim m$ will be changed drastically.

Equation  (\ref{infNq})
can be solved exactly. Note that if we act by the operator $(-{\d^2 \over \d
\te^2} +1)$ on both sides of it we obtain a simple integral equation with the
singular kernel: \be {1 \over \pi} P\int_{-B}^B { d\te' \eps(\te') \over
(\te - \te')^2 }=-h
\label{singeq}
\ee
Its solution is the famous semi-circle law of Dyson:
\be
\eps(\te) = h \sqrt{B^2-\te^2}
\label{semi}
\ee
Its Fourier transform can be expressed through the Bessel function
\be
\eps(\om)= \pi h B J_1(B\om)/\om
\label{epstr}
\ee
Plugging it into (\ref{infNq}) and doing the exact integration over
$\om$ (see \cite{PBM-2}) we obtain the relation (\ref{mh}) between the
$m/h$ and $B$.

Now we can calculate the free energy as a function of $B$ using the
relation (\ref{grst}) and our solution (\ref{semi}):
\be
 f(h) \equiv {1\over N^2}
\Big({\cal E}(h)-{\cal E}(0)\Big) = -{1  \over 8   \pi^2} m h\int_{-B}^B d\te
\cosh \te \sqrt{B^2-\te^2}
\label{free}
\ee
The integral (\ref{free})
together with the relation (\ref{mh})
gives the result (\ref{fh}) for the  energy .

 Let us note that  in the large $N$ limit any virtual and
real processes involve all particles, since   minimal
energies are greater then a minimal separation between masses.
A reasonable
external   field (\ref{hk}) excites all of them on equal footing and leads to
the collective effects. That is why, even though the S-matrix of two,
 say, vector particles
  tends to unity at $N\ra \infty$, the ground
state energy is not that of a free field theory.

\section{Singular behaviour on threshold}

Here we will show that the theory exhibits qualitatively new features
 on the threshold  $h \ra m$ which corresponds
to $B \ra 0$.

At small $B$ asymptotics of the Bessel functions $I_1(B)
\ra 1/\pi(B/2 + B^3/16+...)$ and $K_1(B) \ra 1/B + {B \over 2}
\ln(B/2)+...$. Then, from (\ref{mh}) we obtain:
\be
B^2 \simeq 4 {\Delta \over |\ln \Delta|}, \ \ \ \Delta \ra 0
\label{BD}
\ee
where we introduced $\Delta=h/m-1$. This gives a singular behaviour on the
threshold:
\be
 f(h)  \simeq - (m/2\pi)^2 {\Delta \over |\ln \Delta|},
 \ \ \Delta \ra 0
\label{freD}
\ee

It differs drastically from the threshold behaviour for a finite N theory
of massive  particles, where we would have  3/2 law (see e.g. \cite{W1,JNW}:
\be
 f_N(h) \sim  -m^2 (\Delta)^{3/2}
\label{freN}
\ee

As we already mentioned, the reason for the singular behaviour is the
emergency of
 an extra dimension  in the large $N$ limit - the masses of physical
particles are distanced by each other by the quantity of the order $1/N$, which
is less than any energy scale left in the system. Therefore any external
field excites a bundle of particles - a new extended object, which is
characterized by an extra "momentum" $q$
in addition to the usual momentum $p$.
Let us note that a similar behaviour has been found in the
 quasiclassical limit
of the Sine-Gordon model \cite{Haldane}, \cite{Pokr}.

It is interesting to compare the results (\ref{BD}) and (\ref{freD})
 with the expression for the ground state energy of the $c=1$ matrix
 model \cite{Kaz}. The mechanism by which the inverse logarithmic
behaviour with respect to the cosmological constant occurs also requires
a parametrization  through the fermi level of the corresponding
 fermions (whose coordinates are the
eigenvalues of a hermitean matrix field).
The Fermi level plays the role of a "hidden" parameter
of the problem and the eigenvalues give rise to an extra (Liouville)
degree of freedom of the theory \cite{Das}.

\section{Perturbative regime and the ratio $m/\La_{\bar {MS}}$}

The
field $h$ fixes the scale of energies (with respect to the
mass gap $m$).
 Large $h/m$ regime is the subject of the perturbative theory.
 From eq. (\ref{mh}) we conclude that it corresponds to $B \ra \infty$.
 It follows from the large $B$ asymptotics  of McDonald function $K_1(B)=
\sqrt{{\pi \over 2 B }} e^{-B} \Big(1+{3 \over 8 B}
+ O(1/B^2)  \Big)$ that

\be
{h \over m} =\sqrt{{2 \over \pi}} {e^B \over \sqrt{B}} \Big(1-{3 \over 8 B}
+ O(1/B^2)  \Big)
\label{hmB}
\ee
Solving
 for $B$ we obtain:
\be
B = \ln{h \over m} + {1 \over 2} \ln \ln {h \over m} + {1 \over 2}
\ln{\pi \over 2} + O({1 \over \ln {h \over m}})
\label{Bhm}
\ee
Using the large $B$ asymptotics  $I_1(B)=
\sqrt{{1 \over 2\pi B}} e^B  \Big(1-{3 \over 8 B}
+ O(1/B^2)  \Big)$
we finally
find from (\ref{fh}) and
(\ref{Bhm}):
\be
16 \pi  f(h) = - h^2  B + O({h^2 \over B}) =
- h^2
\Big( \ln{h \over m} + {1 \over 2} \ln \ln {h \over m} + {1 \over 2}
\ln{\pi \over 2} + O({1 \over \ln {h \over m}}) \Big)
\label{bighf}
\ee
This result reproduces correctly one- and two-loop terms of the
perturbation theory as well as
 the universal non-perturbative constant ${1 \over 2} \ln{\pi
\over 2}$ (first calculation of a similar constant was given in
\cite{AndLew} for the Kondo problem).

Let us compare it with the perturbative result \cite{Nied}
 following directly from (\ref{action}). In   our normalizations we have
\be
16\pi  f_{pert}(h) = -8\pi {h^2 \over N {\bar \la}(h)}
\sum_{k=1}^{N}
q_k^2
 - {2 h^2 \over N^2} \sum_{k > j} (q_k-q_j)^2 [\ln |q_k-q_j| -{1 \over
2} ]
 +  O({h^2 \over  \ln (h/m)})
\label{pertf}
\ee
where
\be
q_k= {h_k-h_{k-1} \over  h}, \ \ \  k=1,2,...N; \ \ \ h_0=h_N=0
\ee
and the renormalized coupling in the minimal subtraction (${\bar {MS}}$)
scheme is:
\be
{\bar \la}^{-1}(h) =
{1 \over 4\pi} \Big( \ln{ h \over 2\La_{\bar {MS}}} +
{1 \over 2} \ln \ln { h \over \La_{\bar {MS}}}
+  O(1/ \ln { h \over \La_{\bar {MS}}}   ) \Big)
\label{twol}
\ee
Or,
\be
h {\d \over \d h} {\bar \la}(h) \equiv \beta({\bar \la}) =
-{1 \over 4\pi}  {\bar \la}^2 -
{1 \over 32 \pi^2} {\bar \la}^3
+ O({\bar \la}^4)
\label{betaper}
\ee
Using an approximate solution of the BA equations at finite $N$ for a
particular choice of the
field which
excites  only the vector particles
with the lowest mass $m_1$,
the authors of \cite{Nied}
found the following ratio:
\be
{m \over  \La_{\bar {MS}}} = \sqrt{8\pi/e} {\sin(\pi/N) \over
(\pi/N)} \ra_{N \ra \infty} \sqrt{8\pi/e}
\label{ratio}
\ee
To compare it with our result (\ref{bighf}) we have to calculate all
the sums in (\ref{pertf}) for our choice of fields (\ref{hk}):
\be
q_k ={ \cos{\pi \over N}(k-{1 \over 2}) \over \cos{\pi \over 2 N}}
\ra_{N \ra \infty} \cos x, \ \
0 \le x \le \pi
\label{fields}
\ee
The calculation of corresponding sums in the large $N$ limit
gives:
\be
{1 \over N} \sum_k q_k^2 \ra N \int_0^\pi {dx \over \pi} cos^2x=1/2
\label{sum1}
\ee
\be
\ba{rcl}
&& {1 \over N^2} \sum_{k > j} (q_k-q_j)^2 [\ln|q_k-q_j| -1/2] \ra
 \\
&& \ \\
&&  1/2 \int_0^\pi {dx \over \pi}
\int_0^\pi {dy \over \pi} (\cos x -\cos y)^2 [\ln|\cos x - \cos y| -
{1 \over 2}] = (1 - 2\ln 2)/4    \\
\ea
\label{sum2}
\ee
Now  matching our result (\ref{bighf}) with the perturbation
theory (\ref{pertf}) we obtain the same ratio (\ref{ratio}).
So, two different
choices of the field (that of \cite{Nied} and of our's)
give the same mass gap
in the ${\bar {MS}}$ scheme.

Let us note in the conclusion to this section that we can continue
 comparing  further terms of the expansion (\ref{bighf}) of our exact
ground state energy in the inverse logarithms with the planar part of the
Feynman graph expansion (\ref{pertf}) for the ground state energy
 using  relation (\ref{ratio}).
They must coincide in all orders.

\section{Weak coupling expansion to any order and exact
beta-function}

 The ${\bar \la}(h)$ defined  eq.(\ref{twol}) is equal to  $4\pi /B$ in
the   two-loop approximation (\ref{Bhm}). Since all  next corrections are
nonuniversal we
may
take $4\pi /B$ as the most natural definition of the renormalized
coupling:
\be
 {\bar \la}(h) = 4\pi/B
\label{renla}
\ee

Having in hands the explicit expression (\ref{fh}) for the free energy
we can find all orders of the renormalized perturbation theory. Let us
notice that the function:
\be
y(B)=B K_1(B) I_1(B)
\label{qua}
\ee
satisfies the 3-d order differential equation:
\be
y''' - ({3 \over B^2} +4) y' +{3 \over B^3} y =0
\label{equa}
\ee
If we look for the series expansion of $y(B)$ we find from (\ref{equa})
the recurrence relations on the coefficients, which can be solved.
Finally, by taking  (\ref{renla}) into account we obtain:
\be
 f(h)/h^2 = -{1 \over 4 {\bar \la} } \Big(1-
\sum_{n=1}^{\infty} C_{2n} \Big({{\bar \la} \over 4\pi}\Big)^{2n}\Big)
\label{perex}
\ee
where
\be
C_{2n} = {2n+1 \over 2n-1 } {((2n)!)^3 \over (n!)^4 8^{2n}}
\ra_{n \ra \infty} 2/\sqrt{\pi n} ({ n \over  e})^{2n}
\label{coefs}
\ee
Note that the expansion goes only in even powers.

As we see, in spite of the fact that every coefficient represents a sum
over renormalized planar graphs, it grows factorially with the
order. Most probably this happens because of the renormalons (some
subsequence of logarithmically divergent graphs) giving the main
factorial contribution in each order noticed long time ago by 'tHooft
\cite{tHooft}. This means that we have an exponential number of graphs in
each order but some of them give $(2n)!$ contribution after the momenta
integration.
More than that: the series is a non-signchanging one and thus
non-Borel summable. Nevertheless, the free energy perfectly exists for
any finite ${\bar \lambda}$.
These phenomena seem to be imminent for any asymptotically
free  field theory.

One can still give a prescription how to sum up the series
(\ref{perex}) and
 restore  the result  (\ref{fh}). Notice that:
\be
g(t)=1 - \sum_{n=1}^{\infty} {C_{2n} \over (2n)!}   t^{2 n}
 = F(-1/2,3/2,1,t^2/4)
\label{sum}
\ee
where $F$ is the hypergeometric function.
It has the cut starting from $t^2=  4$.
  We can continue $g(t)$ analytically to the whole
complex plane and make the inverse Borel transform by integrating
 from $0$ to $\infty$ with the exponential factor. The correct
prescription restoring (\ref{fh}) is (see (\ref{renla})):
\be
16 \pi  f(h)/h^2 = - B^2 Re \int_0^\infty dt e^{-t B} g(t)
\label{anali}
\ee

As we see from here   even the Borel nonsummable series in the
asymptotically free theories can be summed up by some
special prescription. This prescription  which uses a
non-trivial procedure of analytical continuation might work well in
other asymptotically free theories. Note that the integral in
(\ref{anali}) taken along the real axes possesses also an
exponentially small ( for $B \ra \infty, \ {\bar \la} \ra 0$) imaginary
part equal to $-K_1^2(B) B^2/\pi$. The similar properties of Borel
integrals in 4d gauge theories were noticed in \cite{Bogom}.

With the definition (\ref{la})
of the running charge one can find from  eq. (\ref{mh}) the exact
beta-function:
\be
\beta({\bar \la}) = h {\d \over \d h}{\bar \la} = - {4\pi \over B^2}
 {\d B \over
\d \ln {h/m} } = - 4 \pi { K_1(B)\over B^2 K_0(B)}
\label{beteq}
\ee
or
\be
\beta({\bar \la}) =
- {1 \over 4\pi} {\bar \la}^2
 { K_1({4\pi \over {\bar \la}})\over K_0({4\pi \over {\bar \la}})}
=- {1 \over 4\pi} {\bar \la}^2
 \sum_{n=2}^{\infty} b_n \Big({{ \bar \la} \over 32 \pi}\Big)^n
\label{betaex}
\ee
where
\be
\ba{rcl}
&&  b_0=1, \ \ b_1=4, \ \ b_2=-8, b_3=64, \ \ b_4= -5^2\cdot 2^5,  \\
&&  b_5=13\cdot 2^{10}, \ \  b_6=-1073 \cdot 2^8, \ \ ,b_7=103 \cdot 2^{16}
, \ \ ...
\ea
\label{coefb}
\ee

\be
b_n \sim  -(-1)^n \sqrt{8/(\pi n)} (4n/e)^n
\label{coefbas}
\ee

\section{Discussion}

Our results could be interesting from several physical points of view:

1.The model considered here is an example
of an exactly
solvable matrix model in 1+1 physical dimensions. A temptation
would be to interprete it as a new string theory in 1+1 dimensional
target space. Recall that a natural interpretation of big planar graphs
could be given
in terms of a sum of world sheets of a
string (see \cite{Kaz} and references therein).
An important condition for it is that the
typical graphs should be big and dense, to describe a smooth surface.
For this one usually tunes the coupling to its (finite) critical value,
corresponding to the "explosion" of the size of graphs. However, in the
asympthotically free theory the critical coupling is equal to zero.
It is the consequence of the fact that the theory has exponential
corrections to the perturbative expansion, leading to the factorial
divergency of its coefficients. In planar theories it can happen because
of the presence of renormalons: some small portion of all the graphs
has a factorially big weight (with respect to the order). These graphs
seem to be not very much suitable for the interpretation in terms of
random surfaces. Even the expansion in terms of the renormalized
coupling contains the same factorial contributions, as we see from our
results. The same scenario should be available to the multicolour QCD,
invalidating this naive relation between planar graphs and world sheets
of a hypothetic QCD string.

2.However, for both PCF and QCD a less trivial string scenario could be
possible. It might exist a (nonperturbative) reformulation in terms of
a new master field, with a new perturbation theory, already suitable
for a string interpretation.
The result
(\ref{freD}) suspiciously reminds the similar result for the 1D
bosonic string (from the matrix quantum mechanics) \cite{Kaz}, if we
take $(h-m)^2$ instead of the cosmological constant $\la-\la_{crit}$.
The
similarity is even more striking if one remembers that for both models
this asymptotics is related to the behaviour of the fermions on the top
of the fermi-sea (\cite{PolWieg,W1}), and the Fermi level is the most
natural physical parameter in both cases. The  semi-circle
law for the distribution of rapidities (37) also suggests that at some
value of field $h$ we deal with an extended object.

3. Another interesting feature of the model (also observed
in the $c=1$  matrix model) is the emergency of an extra dimension
following from the
matrix structure of the theories. This dimension is related to the
random walk along $A_{N-1}$ Dynkin diagram. The structure of this extra
(third) dimension is well seen from the fact that the kernel of the
problem (see eqs.(\ref{kernel}) or (\ref{kerp})) looks as a
propagator of a periodic motion in the space of rapidities versus
 Dynkin diagram. It reminds an extra (eigenvalue) dimension in the
collective field formulation of the 1D
matrix model  usually attributed to the Liouville mode \cite{Das}.

4. One of the immediate problems  is to take into
account
the effects of finite N.
The most interesting effects of finite
$N$'s correspond to the $e^{-c N}$ corrections. To demonstrate it let us look
at the  expansion of the kernel (\ref{kernel}) in the pole terms (see
(\ref{kerp})). The main order for the large N corresponds to the neglection of
all poles but the first (on the finite distance from the origin on
the imaginary axes of $\om$). Next terms will give the exponentially
small contributions (as well as the $1/N$ corrections)
if we treat them as singularities in the
corresponding integral equation.

\bigskip

{\large\bf Acknowledgements}

We would like to thank N.Andrei,
E.Brezin, M.Douglas,
D.Gross,\\
 I.K.Kostov,
A.Neveu, A.A.Migdal, A.M.Polyakov,
A.Sentgupta,
M.Staudacher \\
 and Al.B.Zamolodchikov
for valuable discussions.

\medskip

V.K. is grateful  to the Department of Physics and Astronomy
of the Rutgers University and
to the Mathematical Disciplines Center of the University of
Chicago for the hospitality while this work was in progress.
P.W. was supported
in part by NSF under the Grant DMR 88-19860.

\end{document}